\documentclass{article}
\usepackage{spconf,amsmath,graphicx,hyperref}

\title{An Audio Language Model-Based Voice Concept Bottleneck Framework for Interpretable Health Assessment}
\name{Yu-Wen Chen, Julia Hirschberg}
\address{Department of Computer Science, Columbia University, United States}
\begin{document}
%
\maketitle
\begin{abstract}
Interpretability is critical in clinical decision support. Concept bottleneck frameworks improve it by representing inputs as human-understandable concepts and restricting predictions solely on them. However, research on their use for voice-based health assessment remains limited. In this study, we propose a voice concept bottleneck framework for interpretable health assessment using an audio language model (ALM). The ALM is fine-tuned on a voice quality assessment dataset to enhance its understanding of voice concepts and serves as an independent concept extractor, producing discrete, interpretable scores for a lightweight downstream classifier. The discrete concept scores provide intuitive interpretation, while the lightweight classifier facilitates post-hoc interpretability analyses. Results on depression and dysarthria assessment tasks demonstrate that the proposed framework can flexibly adapt voice concepts to different health conditions and consistently outperforms openSMILE-based and self-supervised speech model-based baselines.

\end{abstract}
\section{Introduction}
\label{sec:intro}

Vocal biomarkers provide valuable insights into an individual's health~\cite{fagherazzi2021voice} and have enabled voice-based approaches for detecting and assessing various health conditions. While end-to-end neural approaches have shown promise~\cite{li2025automated}, their lack of interpretability limits understanding of the factors driving predictions, which is critical for clinical validation and decision-making. One strategy for improving interpretability is to introduce a concept bottleneck, which represents inputs as human-interpretable concepts and restricts predictions to these concepts. In voice-based health assessment, openSMILE~\cite{eyben2010opensmile} is widely employed to extract interpretable low-level acoustic concepts such as pitch, loudness, jitter, and shimmer. However, they are typically combined with less interpretable acoustic features~\cite{al2018detecting, javanmardi2024pre, millet2019learning}, such as MFCCs, limiting the overall interpretability of the resulting models.

Recently, a few studies have explored higher-level voice concepts for health assessment. For example, ~\cite{narain2025voice} investigated the correlation between self-supervised learning (SSL)-based speech model embeddings and interpretable voice-quality dimensions, such as breathiness, in atypical speech. ~\cite{xu2023dysarthria} and ~\cite{leschly2025exploration} developed concept bottleneck models that jointly learn health-related concepts and dysarthria or mild cognitive impairment assessment outcomes. However, jointly optimizing concept learning and downstream prediction may encourage task-specific shortcuts compared with using an independent concept extractor~\cite{margeloiu2021concept}. Moreover, these predefined concept bottlenecks lack flexibility for adaptation across tasks.

Audio language models (ALMs)~\cite{arora2025landscape} can generate flexible, human-understandable descriptions of audio conditioned on input prompts. ALMs have also demonstrated the ability to characterize voice concepts, such as prosody and fluency, associated with speakers' health conditions~\cite{chen2025hearing}. However, existing ALM studies primarily focus on direct downstream applications~\cite{wang2026emotionthinker}. While ALMs are capable of generating reasoning, such reasoning may not faithfully reflect the evidence present in the speech signal and may instead encode task-specific shortcuts. Furthermore, identifying the contribution of each part of the generated reasoning to the model's final prediction is not straightforward. The potential of using ALMs for voice concept bottlenecks with an independent classifier for health assessment has not been explored. 

In this study, we propose an ALM-based voice concept bottleneck framework for interpretable health assessment with three key advantages. First, we mitigate shortcut learning by decoupling the voice concept extractor from the downstream classifier, reducing the risk of task-specific shortcuts. Second, our framework enables flexible concept adaptation. Since diagnostically relevant voice concepts vary across health conditions, our method enables concept adaptation beyond predefined fine-tuning concepts, unlike conventional concept bottleneck approaches. Experimental results on depression and dysarthria assessment tasks show that our approach outperforms openSMILE~\cite{eyben2010opensmile}- and SSL-based baselines, with task-adapted concepts leading to improved predictive performance. Third, our framework enhances interpretability by representing voice concepts on an intuitive 1–5 ordinal scale rather than continuous values. Combined with a lightweight downstream classifier, these designs enable the direct application of post-hoc interpretability methods, such as SHAP~\cite{lundberg2017unified}, to investigate the contribution of each voice concept to final predictions.

\begin{figure*} 
    \centering
    \includegraphics[scale=0.93]{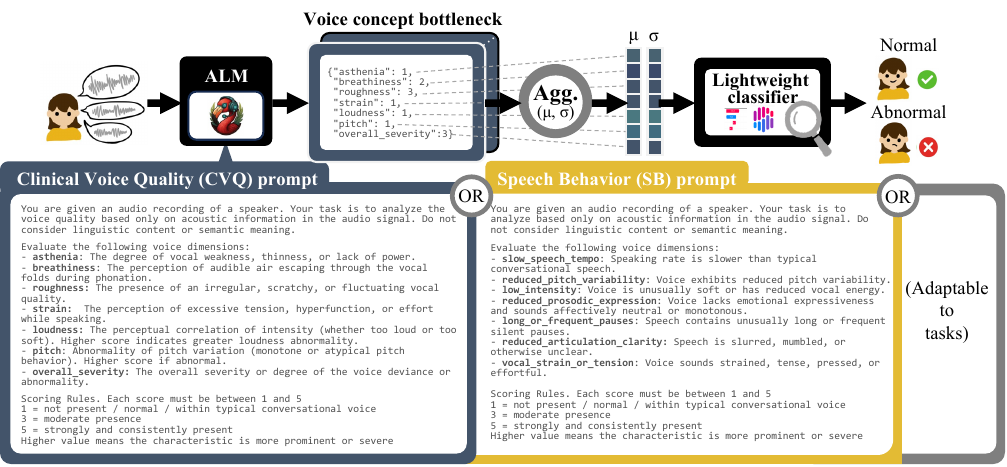}
    \caption{Inference pipeline of the proposed ALM-based voice concept bottleneck framework.}
    \label{fig:proposed}
\end{figure*}

\section{Method}

First, the ALM is fine-tuned on clinical voice quality assessment data to enhance its understanding of medically relevant voice concepts. It then extracts discrete bottleneck concept scores, which are aggregated at the speaker level for health assessment. Figure~\ref{fig:proposed} illustrates the inference pipeline.

\subsection{ALM-based voice concept bottleneck}

To ensure reliable extraction of clinically meaningful concepts, we fine-tune the ALM on a voice quality dataset annotated by experienced clinicians using the Consensus Auditory-Perceptual Evaluation of Voice (CAPE-V) and GRBAS scales. This perceptual supervision guides the model to capture clinically relevant voice concepts. We refer to the integrated attributes from both scales as Clinical Voice Quality (CVQ). During fine-tuning, the ALM is prompted with a CVQ template to optimize its performance on CVQ assessment tasks. In inference, we introduce an additional set of voice concepts, referred to as Speaking Behavior (SB), to evaluate the ALM’s ability to generalize beyond the concepts specified during fine-tuning. While CVQ focuses on clinical voice quality characteristics that are commonly associated with physiological changes in the vocal system, SB emphasizes more on behavioral speech patterns. Detailed descriptions of the CVQ and SB voice concepts are provided in Figure~\ref{fig:proposed}. Voice concepts are measured on a discrete 1–5 ordinal scale, with higher scores indicating a more pronounced vocal characteristic. This ordinal representation is selected because it mirrors human rating conventions and offers a more intuitive interpretation than continuous values.


\subsection{Health assessment with voice concept bottleneck}
We evaluate the proposed framework on two health conditions that affect speaker's voice through distinct clinical mechanisms: depression and dysarthria. While depression is a mental health condition that introduces behavioral alterations, dysarthria is a disorder affecting vocal tract control. This diversity allows us to evaluate the framework's flexibility in adapting its concept bottlenecks to different clinical conditions. We perform speaker-level evaluation by aggregating voice concept scores across all audio clips from the same speaker using their mean ($\mu$) and standard deviation ($\sigma$), forming a speaker-level representation. This representation is used to train a lightweight downstream classifier. Instead of complex neural network architectures, we employ lightweight classifiers to enhance interpretability. By deliberately restricting the capacity of the downstream model, we mitigate the risk that predictions rely on unintended latent shortcuts rather than the intended bottleneck concepts.

\section{Experimental setup}
\subsection{Data}

\textbf{Clinical voice quality}: The Perceptual Voice Qualities Database (PVQD)~\cite{walden2022perceptual} contains recordings of sustained vowels (/a/ and /i/) and six CAPE-V sentences from each speaker, with each speaker annotated with both CAPE-V and GRBAS. For attributes shared by both scales (i.e., overall severity, roughness, breathiness, and strain), we averaged the corresponding scores to obtain unified labels. For attributes unique to a single scale (i.e., asthenia from GRBAS and pitch and loudness from CAPE-V), the original ratings were retained. The dataset was split into training and test sets using an 80:20 speaker-level split. Since the recordings for each speaker are provided as a single audio file, we apply a voice activity detection model\footnote{\url{https://github.com/snakers4/silero-vad}} to segment the audio into individual vowel and sentence recordings. During training, we applied Gaussian noise (amplitude range: 0.001--0.01, $p=0.2$), random gain (gain range: $-2$ to $2$ dB, $p=0.3$), and polarity inversion ($p=0.5$) for augmentation. \textbf{Depression}: We use the interview subset of the Android Corpus~\cite{tao2023androids}, in which all participants respond to the same set of questions posed by an interviewer. The subset contains 116 speakers (52 healthy controls and 64 speakers diagnosed with depression). We use the audio clips provided with the original dataset and follow the official five-fold cross-validation protocol for training and evaluation. \textbf{Dysarthria}: The TORGO corpus~\cite{rudzicz2012torgo} comprises recordings of non-words, isolated words, and both restricted and unrestricted sentences from 8 dysarthric speakers and 7 healthy controls. Only \emph{headMic} recordings are used, with leave-one-speaker-out cross-validation for evaluation.
\subsection{Model settings}

\textbf{Settings for the proposed framework}: For the ALM, we adopt Audio Flamingo Next~\cite{ghosh2026audio} via Hugging Face (\emph{nvidia/audio-flamingo-next-think-hf}). Prior to training, continuous labels are discretized into ordinal scales. We fine-tune ALM via LoRA, configuring all linear layers with a rank of $4$, an alpha of $32$, and a dropout rate of $0.1$. Training is conducted with a learning rate of $1\text{e-}5$ and a batch size of $1$. The model is saved every $100$ steps, and the one yielding the best performance is selected. For the downstream classifier, we utilize an XGBoost model configured with $100$ estimators, a maximum tree depth of $6$, and a learning rate of $0.1$. \textbf{Settings for baseline models}: The baseline methods replace the ALM voice concepts with OpenSMILE or SSL-based features while keeping the rest of the framework unchanged. Specifically, using the same configuration, the downstream classifier takes both mean and standard deviation computed across multiple audio clips from the same speaker as input. For the openSMILE baseline, we use the full \emph{eGeMAPSv02} feature set, although some features in \emph{eGeMAPSv02} are not directly interpretable. For the SSL-based baselines, we adopt Vox-Profile~\cite{feng2025vox}, which uses Whisper as the backbone and extends it with multiple prediction heads to model diverse speech traits. We consider two Vox-Profile variants. (1) Vox-Profile: We employ (\emph{tiantiaf/whisper-large-v3-voice-quality}), which estimates 25 voice quality attributes spanning pitch, texture, volume, clarity, and rhythm dimensions. (2) $\text{Vox-Profile}_{\text{CVQ}}$: We extend the Vox-Profile with additional CVQ prediction heads by fine-tuning it on the PVQD dataset. The model is trained with MSE loss using SGD (lr = $5\text{e-}5$, momentum = $0.7$). Early stopping is applied, and the checkpoint with the best validation performance is retained. For recordings exceeding Vox-Profile's $15$-second inference limit, we apply a $15$-second sliding window with a $7.5$-second hop size. Final representations are obtained by averaging these segment-level predictions over time.

\section{Results}

\subsection{Voice concepts bottleneck for depression assessment}

Table~\ref{tab:performance_depression} presents the performance on depression. Results show that Flamingo fine-tuned with CVQ and inferred with SB (i.e., $\text{Flamingo}_{\text{CVQ} \rightarrow \text{SB}}$) achieves the best performance. Using only 14 discrete features (each with five ordinal levels) with a relatively interpretable classifier (i.e., XGBoost), the model outperforms baselines that rely on less interpretable acoustic features (e.g., MFCCs in BS1~\cite{tao2023androids}, BS2~\cite{tao2023androids}, and OpenSMILE) and architectures (i.e., the LSTM used in BS2~\cite{tao2023androids}). $\text{Flamingo}_{\text{CVQ} \rightarrow \text{CVQ}}$ outperforms $\text{Vox-Profile}_{\text{CVQ}}$, suggesting that ALM-based acoustic bottleneck features are more effective than SSL-based representations. In addition, $\text{Flamingo}_{\text{CVQ} \rightarrow \text{CVQ}}$ outputs discrete scores in 1--5, which are more readily interpretable than the continuous outputs of $\text{Vox-Profile}_{\text{CVQ}}$. The advantage of ALM is further supported by the performance of $\text{Flamingo}_{\text{CVQ} \rightarrow \text{SB}}$. In contrast to SSL-based models, where voice bottlenecks are constrained by fixed heads, ALMs allow flexible redefinition of bottleneck features to better align with downstream tasks. For example, for depression, voice characteristics are more related to speaker behavior than vocal-tract abnormalities, making SB bottleneck features more effective than CVQ.

\begin{table}[htbp!]
\centering
\caption{Performance on depression task. $\text{Flamingo}_{\text{A} \rightarrow \text{B}}$ denotes fine-tuned on \emph{A} and evaluated with the \emph{B} prompt.}
\label{tab:performance_depression}
\renewcommand{\arraystretch}{1.1}
\resizebox{0.85\columnwidth}{!}{
\begin{tabular}{|c|c|c|c|c|c|}
\hline
\multicolumn{1}{|l|}{}                                                  & Precision & Recall & F1    & Accuracy \\ \hline \hline
BS1~\cite{tao2023androids}                                                                     & 0.735     & 0.735  & 0.736 & 0.733    \\ \hline
BS2~\cite{tao2023androids}                                                                     & 0.858     & 0.861  & 0.847 & 0.839    \\ \hline \hline
\begin{tabular}[c]{@{}c@{}}OpenSMILE\\[-3pt] (eGeMAPSv0)\end{tabular}         & 0.752     & 0.773  & 0.752 & 0.725    \\ \hline \hline
Vox-Profile                                                             & 0.682     & 0.755  & 0.703 & 0.657    \\ \hline
$\text{Vox-Profile}_{\text{CVQ}}$                                                         & 0.656     & 0.692  & 0.664 & 0.630    \\ \hline \hline
$\text{Flamingo}_{\text{None} \rightarrow \text{CVQ}}$ & 0.700     & 0.731  & 0.691 & 0.664    \\ \hline
$\text{Flamingo}_{\text{None} \rightarrow \text{SB}}$\  & 0.767     & 0.689  & 0.719 & 0.715    \\ \hline
$\text{Flamingo}_{\text{CVQ} \rightarrow \text{CVQ}}$  & 0.791     & 0.801  & 0.782 & 0.758    \\ \hline
$\text{Flamingo}_{\text{CVQ} \rightarrow \text{SB}}$& \textbf{0.888} & \textbf{0.881} & \textbf{0.871} & \textbf{0.862} \\ \hline \hline
\end{tabular}
}
\end{table}

\begin{figure}[hpbt!]  
    \centering
    \includegraphics[scale=0.84]{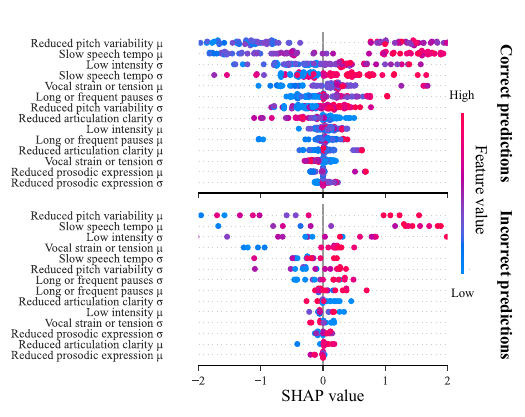}
    \caption{SHAP analysis of depression prediction}
    \label{fig:shap}
\end{figure}

Figure~\ref{fig:shap} shows SHAP~\cite{lundberg2017unified} analysis, which quantifies feature contributions to model predictions for $\text{Flamingo}_{\text{CVQ} \rightarrow \text{SB}}$. For correctly classified samples, reduced pitch variability and slow speech tempo emerge as the two most important features, with their association with depression aligning with psychiatric literature describing depressive speech as slow and monotonous~\cite{cummins2015review}. However, some non-depressed speakers also exhibit reduced pitch variability and slow speech tempo, as these vocal patterns can be influenced by factors other than depression. This overlap highlights the importance of interpretability in automated voice-based health assessment. By enabling clinicians to understand the rationale behind model predictions and assess the supporting evidence, interpretability helps calibrate their confidence in model outputs for clinical decision-making.

\subsection{Voice concepts bottleneck for dysarthria assessment}

\begin{figure}[hpbt!]  
    \centering
    \includegraphics[scale=0.85]{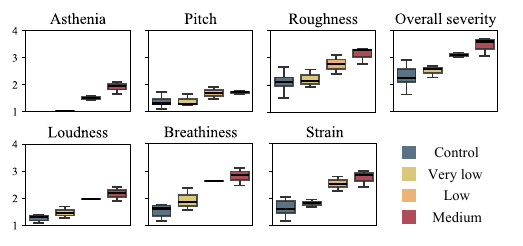}
    \caption{Average CVQ scores across dysarthria groups}
    \label{fig:torgo_pvqd}
\end{figure}

Table~\ref{tab:performance_dysarthria} summarizes the performance on the dysarthria assessment task, including both health status classification (i.e., determining whether a speaker has dysarthria) and severity assessment (i.e., estimating the severity level for speakers with dysarthria). The results demonstrate the effectiveness of the ALM-based acoustic bottleneck approach, with $\text{Flamingo}_{\text{CVQ} \rightarrow \text{CVQ}}$ achieving the best overall performance. In contrast to the depression assessment task, the CVQ feature set outperforms the SB feature set because dysarthria directly impairs the physiological mechanisms of speech production, making clinical voice quality a more direct indicator. Figure~\ref{fig:torgo_pvqd} shows box-plots of the average CVQ bottleneck feature scores across different groups extracted by $\text{Flamingo}_{\text{CVQ} \rightarrow \text{CVQ}}$. The results are consistent with the intuition that clinical voice quality abnormalities increase with dysarthria severity. Notably, for acoustic features such as asthenia, pitch, and roughness, the control and very low severity groups exhibit highly similar distributions, while a more pronounced separation emerges between the very low and low severity groups.

\begin{table}[htbp!]
\centering
\caption{Performance on dysarthria task}
\label{tab:performance_dysarthria}
\renewcommand{\arraystretch}{1.1}
\resizebox{0.85\columnwidth}{!}{

\begin{tabular}{|c|c|c|c|}
\hline
    & ACC-class & ACC-level & PCC-level \\ \hline \hline
Hubert-24+SVM~\cite{javanmardi2024pre}                                                         & 0.761     & -         & -         \\ \hline
Hubert-10+CNN~\cite{javanmardi2024pre}                                                         & -         & 0.498     & -         \\ \hline
eGeMAPS+SVM~\cite{javanmardi2024pre}                                                          & 0.594     & 0.421     & -         \\ \hline \hline
\begin{tabular}[c]{@{}c@{}}OpenSMILE\\[-3pt] (eGeMAPSv02)\end{tabular}      & 0.533     & 0.375     & 0.348     \\ \hline \hline
Vox-Profile                                                           & 0.800     & 0.500     & 0.467     \\ \hline
$\text{Vox-Profile}_{\text{CVQ}}$                                                       & 0.800     & 0.625     & 0.577     \\ \hline \hline
$\text{Flamingo}_{\text{CVQ} \rightarrow \text{CVQ}}$ & \textbf{0.800} & \textbf{0.750} & \textbf{0.894} \\ \hline
$\text{Flamingo}_{\text{CVQ} \rightarrow \text{SB}}$ & 0.733     & 0.375     & 0.554     \\ \hline \hline
\end{tabular}
}
\end{table}

\subsection{Performance on voice concepts learning}

We calculate the Pearson correlations ($r$) between the CVQ bottleneck scores and human ratings. Before fine-tuning, the ALM's ($\text{Flamingo}_{\text{None}}$) scores for asthenia ($r$\texttt{=}-0.018), abnormal loudness ($r$\texttt{=}-0.019), abnormal pitch ($r$\texttt{=}-0.008), and roughness ($r$\texttt{=}-0.005) show little to no correlation with human ratings (average $r$\texttt{=}0.132). The poor correlation for abnormal loudness and pitch probably stems from a mismatch between the ALM’s pretraining bias and CVQ definitions: ALM tends to associate higher pitch or loudness with higher scores, whereas CVQ ratings assign higher scores to both \emph{abnormally} high and low levels. In contrast, overall severity shows the highest correlation among all concepts ($r$\texttt{=}0.482), perhaps because the model infers impairment from semantic coherence in the recordings. After fine-tuning, the SSL-based ($\text{Vox-Profile}_{\text{CVQ}}$) and the ALM-based models ($\text{Flamingo}_{\text{CVQ}}$) showed comparable performance on CVQ assessment, with moderate correlations with human ratings on average ($\text{Vox-Profile}_{\text{CVQ}}$ $r$\texttt{=}0.562, $\text{Flamingo}_{\text{CVQ}}$ $r$\texttt{=}0.519) .

\section{Conclusion}
We introduce an ALM-based voice concept bottleneck framework for interpretable health assessment. We focus on interpretable voice concepts that go beyond commonly used low-level acoustic features (e.g., absolute pitch and loudness) by incorporating information from clinical voice quality. Our findings reveal that the pretrained ALM fails to faithfully capture these voice concepts, establishing a clear need for fine-tuning. Crucially, this fine-tuning does not restrict the ALM to supervised concepts; instead, it enhances the model's general capability in voice concept extraction, resulting in better performance on downstream tasks through adaptable concepts. Overall, we demonstrate the potential of ALMs for more interpretable clinical decision-support systems. 



\bibliographystyle{IEEEbib}
\bibliography{strings,refs}

\end{document}